\documentclass[10pt]{article}
\usepackage[margin=1in]{geometry}
\usepackage{subcaption}
\usepackage{graphicx}

\usepackage{times}

\usepackage{url}
\usepackage{hyperref}
\usepackage{subcaption} %Nirav commented
\usepackage{graphicx}
\usepackage{amsmath}
\newcommand{\citepos}[1]{\citeauthor{#1}'s (\citeyear{#1})}

\usepackage[numbers,square,sort]{natbib}
\usepackage{booktabs}

\newtheorem{example}{Example}

\usepackage{placeins} %Nirav commented
\usepackage{dcolumn}
\usepackage{tikz}
\usetikzlibrary{shapes,arrows,calc,fit,shadows,backgrounds,automata,positioning}
\usepackage{pgfplots}
\pgfplotsset{compat=newest}
\usetikzlibrary{pgfplots.statistics}
\usepackage[inline]{enumitem}
\setlist{noitemsep}
\usepackage{listings} %Nirav commented

\newcommand{\mycaption}[1]{\caption[#1]{#1.}}

\usepackage{xspace}

\newcommand{\fbf}{\textbf}

\newcommand{\fsl}{\textsl}

\newcommand{\fsub}{\textsubscript}

\usepackage{balance}

\usepackage[]{xcolor}

\title{Norms and Sanctions as a Basis for Promoting Cybersecurity Practices}

\author{Nirav Ajmeri$^\dagger$, Shubham Goyal$^\ddagger$, and Munindar P.~Singh$^\dagger$\\
$^\dagger$North Carolina State University, Raleigh, NC, USA \\
$^\ddagger$Amazon, Seattle, WA, USA \\
najmeri@ncsu.edu, gsshubha@ncsu.edu, mpsingh@ncsu.edu
}
\date{}

\begin{document}
\sloppy
\twocolumn

\maketitle
\pagenumbering{arabic}
\thispagestyle{plain}
\pagestyle{plain}

\begin{abstract}
Many cybersecurity breaches occur due to users not following good cybersecurity practices, chief among them being regulations for applying software patches to operating systems, updating applications, and maintaining strong passwords. 

We capture cybersecurity expectations on users as \emph{norms}. We empirically investigate sanctioning mechanisms in promoting compliance with those norms as well as the detrimental effect of sanctions on the ability of users to complete their work. We realize these ideas in a game that emulates the decision making of workers in a research lab. 

Through a human-subject study, we find that whereas individual sanctions are more effective than group sanctions in achieving compliance and less detrimental on the ability of users to complete their work, individual sanctions offer significantly lower resilience especially for organizations comprising risk seekers. 
Our findings have implications for workforce training in cybersecurity.
\end{abstract}

\section{Introduction}

As computing has spread into every part of our economic and personal lives, two related trends are apparent: (1) cybersecurity threats can place more and more of our welfare at risk; and (2) attackers have more to gain from successful attacks and, therefore, the number and variety of attacks are proliferating. At the same time, user behavior inadvertently provides pathways through which attackers can succeed.  

Recognizing the challenges in user behavior, organizations have created cybersecurity regulations \cite{Such+19:cyber-hygiene} to codify best practices such as keeping operating systems and applications up to date, disabling unused network ports and services, and not sharing passwords. Following the hygiene metaphor, we adopt the term \emph{immunity tasks} for such practices. However, compliance with regulations is rarely adequate and security breaches continue to occur.  As an illustration, \citet{ICSE-17:Semaver} documented healthcare security and privacy breaches (each of which affected 500 or more patients) with respect to user behavior in light of security and privacy regulations.

Because the actions or nonactions (e.g., carelessness) of one user affect outcomes for others, achieving good cybersecurity practices maps to social norms. 
Sanctions \citep{Nardin+16:Sanctioning} provide a recognized means to establish norms but have not been studied in connection with cybersecurity. 

%% However, there is little research exploring what mechanisms could reduce the adverse influence of human factors in system security.

\paragraph{Contributions.}
We investigate how sanctions can promote cybersecurity practices. Specifically, we investigate three research questions in reference to sanction types (group, individual, and peer, which we explain below) and risk attitudes.

\begin{itemize}
\item How effectively does a sanction type lead to improved cybersecurity practices?
\item How detrimental is a sanction type to user productivity?
\item What influence does a sanction type has on workers with different risk attitudes?
\end{itemize}

\paragraph{Approach and findings in brief.}
We develop a game to simulate a real-life work setting, such as a corporate office, where workers complete assigned tasks while using computers. Each player assumes the role of an office worker and is challenged to complete assigned tasks (captured as points earned) along with maintaining the security of their computer. Failure to complete an immunity task may attract sanctions, causing a loss in points earned or the loss of an opportunity to earn points.

Our human-subject study using this game yields the following findings. First, workers complete more tasks and are sanctioned less often under individual sanctions than under group sanctions. Experience improves the advantage for individual sanctions in promoting compliance. 
Second, workers under individual sanctioning are more productive and lose less time due to being sanctioned than under group sanctioning.
Third, individual sanctioning yields lower resilience with risk-seeking workers, i.e., risk seekers recover slower from an attack than risk-averse workers.

In addition,  group sanctioning leads to twice the number of peer sanctions than individual sanctioning, which indicates the potential for a self-regulating environment in which users monitor each other and intervene to improve adoption of cybersecurity practices. Group sanctioning also yields better resilience with workers having risk-seeking attitudes.

\section{Related Work}
\label{sec:related}

We describe selected research most relevant to this work.

\citet{Beris+15:security-risk} study workers in organizations to understand their risk attitudes regarding cybersecurity and how those attitudes correlate with security practices.  They find that compliance is improved when security policies are aligned with productive activity. In contrast, we seek to understand how users behave with the end of identifying interventions, including by users themselves, that can improve the adoption of good security practices.

\citet{Frey+19:security-game} relate cybersecurity with gameplay as a basis for exploring security decision making by stakeholders, e.g., where to make investments in the way of infrastructure. Their game considers external threats. In contrast, our study is focused on interventions that improve compliance while maintaining productivity.

We draw upon research on norms outside of cybersecurity.
We adopt the idea of sanctions as negative rewards from research on norms \citep{Aldewereld-TAAS16-GroupNorms,Andrighetto-2013-PunishVoice,noussair2005combining}.
Importantly, sanctions are valuable in helping an agent learn applicable norms by experiencing positive or negative sanctions applied as a result of their actions. 

\citet{Mahmoud-AAMAS2016} propose a resource-aware adaptive punishment technique that enables norm establishment with larger neighborhood sizes than resource-unaware punishment. They evaluate the adaptive technique via a simulation. 
\citet{DellAnna-AAMAS19-Runtime} develop a norm revision mechanism that considers agents' preferences and revises norms by revising associated sanctions at runtime. They evaluate their mechanism on a traffic simulator ring road environment. 
% and find that their norm revision mechanism identifies optimal sanctions. 

\citet{Andrighetto-2013-PunishVoice} suggest combining punishments (negative sanctions) with norm communication to improve compliance, instead of using the two separately. They conduct a human-subject study where subjects play a standard public-goods game to evaluate their hypothesis. 

\citet{Patel+16:expectations} study how people hold expectations about each other's behavior and how the violation of such expectations influences their interactions with each other. They focus on software engineering and analyze mobile app features and reviews in their study. 
\citet{Goyal-AAMAS19-Hygiene} study cybersecurity compliance in a game environment. They, however, do not investigate intricacies such as how workers' risk attitudes influence norm compliance---which we investigate here. 

\citet{villatoro2014norm} study the effectiveness of group sanctions in enforcing norm compliance. They find group sanctions as a powerful way to regulate behavior via a laboratory experiment. 
\citet{Aldewereld-TAAS16-GroupNorms} show how group norms, which apply to a group, differ from individual norms in terms of responsibility and fulfillment. Norms can be internally enforced, e.g., when noncompliance triggers emotions of shame or guilt. Such internal controls are valuable in settings where it may be difficult to monitor and enforce compliance. We study group sanctions since they may lead to improved compliance.

\section{Examples: Individual, Group, Peer Sanctions}
\label{sec:Examples}

Individual and group sanctions are applied by a manager, a designated party who has the responsibility of monitoring and sanctioning, and peer sanctions are applied by users.

\begin{example}[Secure Co.]
Consider Alex, Bob, Charlie, and Dave, who are software developers working at Secure Co. Each of them has a workstation connected to the same network. Each developer is tasked with one or more projects using tools available on his or her workstation. Each software developer has a risk attitude and other personality traits. Additionally, Secure Co. has defined cybersecurity regulations such as updating passwords periodically, and installing security patches. The developers are expected to comply with these regulations. Erin is an IT manager who looks after compliance with cybersecurity regulations.
\end{example}

Consider a case where Alex does not patch his workstation in time against an OS vulnerability. Erin observes that Alex's workstation was not patched. She disconnects it from the local network so that other workstations are not at risk, and patches it. Erin takes some time to patch the computer, during which Alex cannot work. Erin's disconnecting Alex's workstation is an example of an \emph{individual sanction} where the person who failed to comply with the regulation is sanctioned \citep{Nardin+16:Sanctioning}. 

Alternatively, on noticing that Alex's workstation was not patched in time and could have affected other workstations on the same network, Erin along with disconnecting Alex's workstation, disconnects Bob's, Charlie's, and Dave's workstations. Erin disconnecting all the workstations prevents all developers from working on their respective tasks. This is an example of a \emph{group sanction}. 

Consider another case where Alex has not patched his workstation. Bob notices that Alex has not patched his workstation and fears that an exploit of a vulnerability on Alex's workstation could result in all their workstations being compromised. Bob frowns at Alex and asks Alex to patch his workstation, suggesting that all the workstations on the network would be at risk. Bob's frowning is an example of a \emph{peer sanction} \citep{Nardin+16:Sanctioning,haynes2017engineering}.

\section{Model of Enterprise Cybersecurity}
\label{sec:game-model}
We now delineate a simplified multiagent systems model of work in an enterprise and how it relates to cybersecurity. 

\noindent
\fbf{Active entities.}
This model has three active entities. 
For simplicity, we assume a unique \emph{manager} and a set of two or more \emph{workers}.  Each worker uses a computer and administers some aspects of the computer, such as the timing of any software updates. Attackers seek to compromise the workers' computers. We don't model attackers explicitly but include them in the \emph{environment}---the environment may launch cyberattacks against the enterprise.

\noindent
\fbf{Tasks.}
A worker is assigned projects, which comprise \emph{project tasks} that the worker must complete. In addition, a worker performs \emph{immunity tasks}, such as patching the operating system, upgrading the firewall, and changing a password to maintain compliance with security policies.

\noindent
\fbf{Norms and sanctions.}
A \emph{norm} characterizes the correct interactions between agents. Specifically, a commitment is a type of norm that captures what a worker is expected to do under what circumstances and a prohibition is a type of norm that captures what a worker is expected to refrain from doing under what circumstances. We model norms that apply on workers with respect to each other or the manager, but not on the manager or the environment. A worker is \emph{accountable} for each applicable norm. When a worker does not follow a norm for which he or she is accountable, we state that the norm is violated by that worker. 

We model enterprise security policies as well as workers' expectations of each other as norms. Norms that apply to each worker with respect to the manager and each other are to patch his or her computer and update his or her password. For simplicity, we write these norms as Patching and Updating Passwords, respectively. Suppose Secure Co. requires workers to update passwords every month: hence, a commitment applying on each worker. A worker who has not updated password for more than a month is in violation.

A \emph{sanction} is a response to a norm violation. We adopt ideas from \citepos{Nardin+16:Sanctioning} typology. For simplicity, we consider only negative sanctions (punishments).
In real life, completing an immunity task does not provide explicit rewards but not completing it can lead to negative consequences in the form of cyberattacks. In the same spirit, we adopt negative sanctions as a consequence of not following a norm.

We consider three types of sanctions. 

\begin{itemize}
\item An \emph{individual} sanction is applied by the manager to a specific worker. The manager can partially observe the security of all computers and can determine if a norm violation has occurred. For example, a manager may disconnect a worker's computer from the enterprise network until the worker addresses the security weakness.

\item A \emph{group} sanction is applied by the manager to the entire set of workers. For example, a manager may disconnect all workers' computers from the enterprise network until the one worker who violated security policies addresses the security weakness on his or her computer. That is, other workers would suffer for no fault of theirs.

\item A \emph{peer} sanction is a sanction from a worker to a peer. Each worker can observe a peer's compliance with norms. The wish to avoid a group sanction by manager can motivate a compliant worker to peer sanction a noncompliant worker.
\end{itemize}

\noindent
\fbf{Actions.}
Each worker may choose an action from available alternatives given its resources and beliefs. A worker can perform the following actions: (1) perform a project task; (2) perform an immunity task; (3) observe and peer sanction a colleague; and (4) skip---when the worker has just been sanctioned by the manager or a peer. A peer sanction is automatically lifted after one time step.

The manager observes the system state and sanctions a worker. The manager imposes sanctions probabilistically upon norm violation---to capture the idea that a noncompliant worker can escape notice and may yet suffer an attack. Each sanction is lifted after two time steps. 

The environment's action is to attack.

\noindent
\fbf{States.}
The system state composes the states of the workers and their computers. A worker's state is a pair of his or her compliance status (whether compliant or noncompliant) and sanctioning status (whether currently sanctioned or not). Initially, a worker is compliant and not sanctioned.

A worker transitions between these states depending upon his or her actions and those of the manager and of other workers, and of the environment.
\begin{itemize}
    \item A project task is enabled except when the worker is in a sanctioned state and doesn't affect either the compliance or the sanctioning components of its state.
    \item An immunity task is enabled only when the worker is in the noncompliant state and restores compliance (once all security vulnerabilities are addressed) but doesn't affect the sanctioning component.
    \item A peer sanction by the worker on someone else doesn't affect either component.
    \item A peer sanction received by the worker doesn't affect compliance status but transitions into a sanctioned state. The worker who issues the sanction loses a step.
    \item An individual or group sanction received by the worker doesn't affect compliance status but transitions into a sanctioned state.

\end{itemize}

A computer cycles between three states (safe, vulnerable, and unusable).
Initially, a computer is safe. A computer transitions to the vulnerable state when there is an attack. The worker assigned to the computer must perform an immunity task to restore it to a safe state. However, the worker may choose not to perform the immunity task and instead continue with a project task. An attack on a safe computer makes it vulnerable (modeled as immunity loss) but an attack on a vulnerable computer causes it to become unusable. When that happens, the worker loses all incomplete work and must perform an immunity task to restore the computer to the safe state before proceeding.

\subsection{Game Design}
\label{sec:game-design}

\begin{figure*}[!htb]
\centering
\input{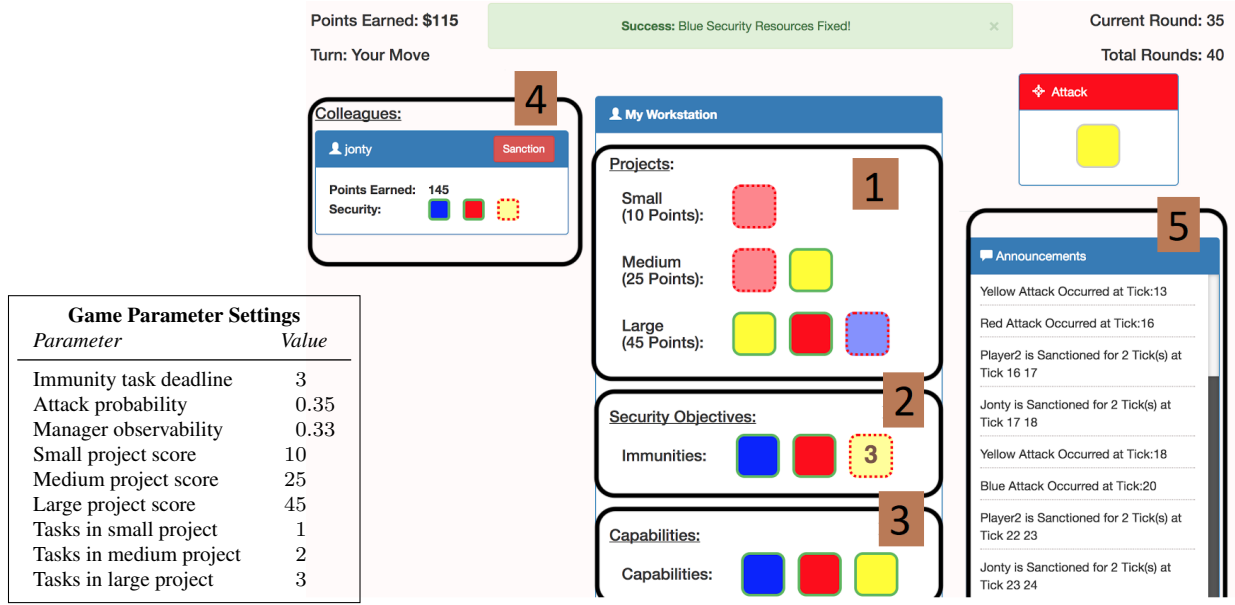}
\caption{Game screenshot. Dashboard view of a single player showing five parts of the game. (Parameter settings are overlaid on the left.)} 
\label{fig:screen}
\end{figure*}

A game enables learners to experience situations that are infeasible in the real world for reasons of safety, cost, and time.  
Role-playing is an established technique for educational activities, and can foster intrinsic motivation. 
Also, users are familiar with rewards via reputation points, badges, and leader boards. We designed a web-based game, a screenshot of which is shown in Figure~\ref{fig:screen}, following the above model. 
% \footnote{We will make the code available on acceptance.}
% We have made available the code base on \mps{let's post to either an NCSU github account for the lab or on the lab page} GitHub.\footnote{\url{https://github.com/shubham2892/Sanction-security-game}}\mps{remove for blinding} 

The game considers three tasks. $T_{\mathrm{blue}}$, $T_{\mathrm{red}}$ and $T_{\mathrm{yellow}}$, represented by blue, red, and yellow tiles. Part~1 of Figure~\ref{fig:screen} shows the project section. There are three types of project, small, medium, and large, with one, two, and three tasks, respectively. The set of tasks that makes up a project is chosen randomly. For instance, in Figure~\ref{fig:screen} the medium project is made up of red and yellow tasks. After a player completes all the tasks of a project, new resources are assigned to the project randomly. A player needs to complete all the tasks in a project to gain the points associated with the project. 

Each task is mapped to immunity ($I_{\mathrm{i}}$) and capability ($C_{\mathrm{i}}$) shown in Parts~2 and~3, respectively. Each attack is directed toward a particular immunity. After an attack $A_{\mathrm{i}}$, the player loses immunity $I_{\mathrm{i}}$. For instance, a yellow attack takes away the yellow immunity. This is equivalent to a computer being in the vulnerable state. The player is given a deadline $D_{\mathrm{i}}$. The player can transition back to the safe state by performing the requisite immunity task---in game terms, clicking relevant tiles in the immunity section. If the player does not perform the immunity task by deadline $D_{\mathrm{i}}$, the player transitions to the noncompliant state and can be sanctioned by the manager. (The manager is not a player but is automated in the game engine.)

If there is an attack $A_{\mathrm{i}}$ for which the immunity $I_{\mathrm{i}}$ was already lost, the player loses the associated capability $C_{\mathrm{i}}$. Losing a capability is equivalent to the computer being in unusable. When a player loses the capability $C_{\mathrm{i}}$, the player cannot complete the corresponding task $T_{\mathrm{i}}$ until the player completes the associated immunity tasks. 

A player's objective in the game is to earn the most points analogous to compensation from completing projects. Players can see each other's current score (Part~4). 

The game is divided into 40 rounds: each player must make a move in each round or explicitly skip it.

\subsection{Realizing our Enterprise Model in the Game}

Each component of our enterprise model maps directly to the game. The players are workers; the manager and environment are implemented in the engine.

Figure~\ref{fig:screen} (left) shows that an immunity task has a deadline of three rounds, i.e., the player gets three rounds to complete the immunity task after the immunity is lost without the risk of being sanctioned.
A norm on a player is to complete the immunity task before the deadline.

\noindent
\fbf{Player Actions.}
A player can take one of these actions:

\begin{itemize}
\item \emph{Complete project task.}  Figure~\ref{fig:screen} (left) shows the points awarded and the number of tasks required for each project type. A player is awarded a score corresponding to a project only after he or she completes all of that project's tasks.

\item \emph{Complete immunity tasks.} A player does not receive points for completing an immunity task but should complete it to avoid sanctions and to retain availability of a resource. 

\item \emph{Peer sanction.} A player can peer sanction another player.

\item \emph{Skip.} When a player is sanctioned by the manager, the player must skip that turn and clicks the pass button to acknowledge the loss of a turn.
\end{itemize}

\noindent
\fbf{Manager Actions.}
The manager (with some probability) observes the immunities of each player at the beginning of each round. If the manager observes that a player has not completed an immunity task past its deadline, the manager sanctions the player at the beginning of the round.
% with a probability equal to the number of incomplete immunity tasks multiplied by the manager's probability of observing a violation. 
% 
% \mps{confusing; I changed it based on your previous comment from "manager's observability" to "manager's probability of observing a violation" but i don't see why the probability would be multiplied in since that would be implicit in when the sanction is applied---double counting probability; if anything it should be divided by probability to penalize the violator if the monitoring occurs at low probability} \nsa{``multiplying count of incomplete tasks with probability of observing'' is not required. Removing it}
The sanction can be group or individual, depending on the game. 

\noindent
\fbf{Environment Actions.}
We set the attack probability to 0.35 for all games, as listed in Figure~\ref{fig:screen} (left), so that the attacks in a game are neither too many nor too few. Too high an attack probability would result in players losing their capabilities frequently and too low an attack probability would lead them to suffer no loss. The former would induce excessive caution (always complete immunity tasks first) and sloppiness  (always complete project tasks first), respectively. We wanted the players to constantly choose between the immunity task and project tasks. 

There are four types of attack. Three of them, $A_i$ (for $i$ being blue, red, or yellow, result in either the loss of the corresponding immunity, $I_i$, or if the immunity is already lost, then in the loss of availability of the corresponding resource $R_i$. $A_{\mathrm{black}}$ attack is equivalent to the other three attacks happening simultaneously. A player must complete all immunity tasks within their respective deadlines. The probabilities of blue, red, and yellow attacks are equal and three times the probability of the black attack.

\noindent
\fbf{Sanctions.}
There are two ways of sanctioning in the game. One, via a group or individual sanction by the manager, where a player is forced to pass a certain number of rounds.  The number of rounds skipped, $n$, is equal to twice the number of unfinished immunity tasks. After $n$ rounds, the player gains back the immunity for which he or she was sanctioned. 

Two, via loss of a capability needed for a task $T_{\mathrm{i}}$. In this case, the player can no longer complete a task until the immunity is regained. For example, if a player loses the red capability after two red attacks, the player cannot complete the red task until the player completes the red immunity task.

\section{Experimental Setup}

Our study included the following steps.

\begin{description}

\item[Risk Attitude Survey.] 
We employ DOSPERT to assess participants' risk-seeking attitude \citep{weber2002domain}. DOSPERT is a psychometric scale to assess risk seeking in five content domains: financial decisions, health and safety, recreational, ethical, and social decisions.
A higher score on DOSPERT scale indicates a greater willingness to take risks.
We choose DOSPERT because of its compactness. 
Longer alternatives would leave participants with less time for gameplay.

\item[Training.] Each participant watched a five-minute video explaining the game, and then played two demo games, excluded in our statistical evaluation. Participants were informed that these games will not be evaluated and were encouraged to familiarize themselves with different elements of the game. Demo games demonstrated individual and group sanction.

\item[Gameplay.] Each participant played four games. Two with individual sanctions (InS$_1$, InS$_2$) and two with group sanctions (GrS$_1$, GrS$_2$), in arbitrary order. After each game, participants completed a short post-game survey to record their opinions on the sanction policy employed in the game.

We configure each game to apply either group or individual sanctions in addition to peer sanction.

\item[Post Survey.] After the gameplay, participants completed a post-study survey capturing their overall feedback.  

\item[Incentives.]
Each player received a payment that had a fixed component as well as a bonus based on their score. The motivation for the bonus was to encourage players to perform well.

\end{description}

\section{Metrics}
We computed the following metrics. 
\begin{description}

\item[Compliance] measures the frequency of a worker being compliant. We measure compliance via these metrics:

\begin{itemize}
\item \emph{Completed Immunity Tasks:}  After an attack, a player loses an immunity ($S_{\mathrm{i}}$) and is given a deadline ($D_{\mathrm{i}}$) to fix it. Regaining immunity is through the completion of an immunity task.

\item \emph{Manager Sanctions} counts the number of sanctions by the manager, which happen when a player does not complete an immunity task by its deadline.
For individual sanction, we calculate the total number of sanctions issued. 
For group sanction, we calculate the total number of sanctions issued to the group, irrespective of the number of players in the group.
\end{itemize}

\item[Efficacy] measures how productive participants are in completing their project. We measure efficacy via these metrics:

\begin{itemize}

\item \emph{Score:} Cumulative value gains from the tasks completed, indicating productivity.

\item \emph{Rounds skipped:}  Whenever a player is sanctioned by peers or the manager, the player is forced to skip one or more rounds, which indicates loss in productivity due to noncompliance.

\end{itemize}

\item[Resilience] measures how quickly the system returns to the state of being norm compliant. We measure resilience by calculating the number of rounds taken by a player to regain an immunity after losing it.

\end{description}

\section{Results and Discussion}
\label{sec:result}  
The study sought to compare how people respond to each sanctioning mechanism and how their productivity in completing project task is affected by these mechanisms. 

We conducted our study on Amazon Mechanical Turk (MTurk) (available at \url{https://www.mturk.com/mturk/})
% \citep{Turk-11} 
where participants played the game. 
The study was approved by our university's Institutional Review Board. We collected an informed consent from each participant and provided a payment on completion of the study.
Participants were allowed to participate only once in the study to mitigate the threat of learning. The study was conducted in slots of 60 minutes. Thirty participants participated in the study playing 107 games. The group size for the games varied from two to five.

All game parameters other than the sanctioning method were fixed throughout the study (Figure~\ref{fig:screen}). We recorded every move made by a player in every game and evaluated this data to compare group and individual sanction with respect to each of the measures described below. 

We test significance via the two-tailed paired \emph{t}-test. We measure the effect size using Hedges' $g$ \cite{Cohen-88:Statistics}, which is computed as the difference in the means divided by the pooled weighted standard deviation. We choose Hedges' $g$ to measure effect size because it is better suited for smaller sample sizes. Recognizing some caveats, we adopt \citepos{Cohen-88:Statistics} suggestion to interpret effects above 0.20, 0.50, and 0.80 as small, medium, and large.

\subsection{Compliance}
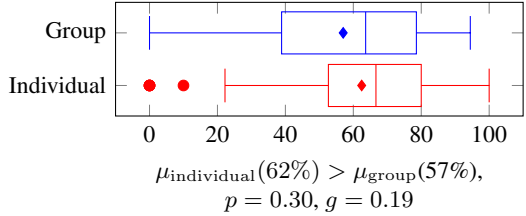
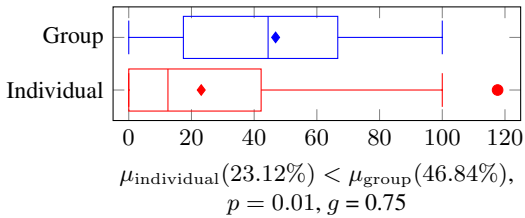
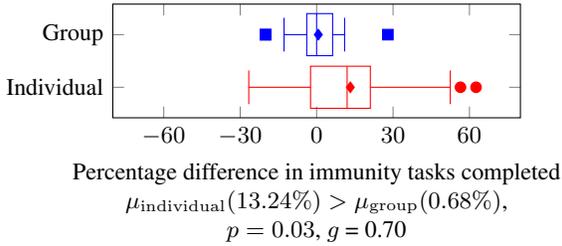
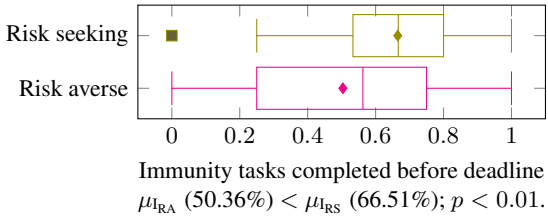
\begin{figure}
\centering
% effective size - 0.205962
\begin{subfigure}[t]{0.90\columnwidth}
  \centering
  \begin{tikzpicture}
    \tikzstyle{every node}=[font=\small]
    \begin{axis}[
	y=0.7cm,
	ytick={1,2},
	yticklabel style={align=center},
	yticklabels={Individual,Group},
	width=7cm,
	xtick={0,.2,.4,.6,.8,1},
	xticklabels={0,20,40,60,80,100},
	xlabel={$\mu_{\mathrm{individual}}(62\%)>\mu_{\mathrm{group}}$(57\%),\\$p=0.30$, $g= 0.19$},
	xlabel style={align=center},
	xmin=-0.1, xmax=1.1,
	boxplot/average=auto,
	title style={align=center},
	title={},
	title style={yshift=-1ex,},
	]
	\addplot+[red,boxplot,mark options={fill=red}]
	table [row sep=\\,y index=0] {
	data\\
    0.8\\ 1\\ 0.8\\ 1\\ 1\\ 0.8\\ 0.92\\ 0.76\\ 0.32\\ 0.52\\ 0.6666666666666666\\ 0.75\\ 0.9166666666666666\\ 0.5833333333333334\\ 0.75\\ 0.75\\ 0.75\\ 0.65\\ 0\\ 0.9\\ 0.1\\ 0.8\\ 0.5625\\ 0.625\\ 0.5625\\ 0.25\\ 0.5833333333333334\\ 0.25\\ 0.7333333333333333\\ 0.7333333333333333\\ 0\\ 0.6666666666666666\\ 0.8666666666666667\\ 0.6666666666666666\\ 0\\ 0.5333333333333333\\ 0.8\\ 0.9\\ 0.5333333333333333\\ 0.8666666666666667\\ 0.9375\\ 0.5\\ 0.8095238095238095\\ 0.6190476190476191\\ 0\\ 0.7142857142857143\\ 0.42857142857142855\\ 0.2222222222222222\\ 0.6111111111111112\\ 0.7222222222222222\\
	};
	\addplot+[blue,boxplot,mark options={fill=blue}]
	table [row sep=\\,y index=0] {
	data\\
    0.3888888888888889\\ 0\\ 0.5\\ 0.3888888888888889\\ 0.3888888888888889\\ 0.6363636363636364\\ 0.8181818181818182\\ 0.7272727272727273\\ 0.6363636363636364\\ 0.6363636363636364\\ 0.7692307692307693\\ 0.9230769230769231\\ 0.7692307692307693\\ 0.6923076923076923\\ 0.8461538461538461\\ 0.6363636363636364\\ 0.6363636363636364\\ 0.8181818181818182\\ 0.9444444444444444\\ 0.3333333333333333\\ 0.8333333333333334\\ 0\\ 0.4666666666666667\\ 0.26666666666666666\\ 0.5333333333333333\\ 0\\ 0.75\\ 0.3333333333333333\\ 0.6666666666666666\\ 0.25\\ 0.4166666666666667\\ 0.5\\ 0.3333333333333333\\ 0.6\\ 0.5333333333333333\\ 0.6666666666666666\\ 0.5833333333333334\\ 0\\ 0.5\\ 0.4444444444444444\\ 0.3888888888888889\\ 0\\ 0.3333333333333333\\ 0.7857142857142857\\ 0.7857142857142857\\ 0.8333333333333334\\ 0.8333333333333334\\ 0.8666666666666667\\ 0.8\\ 0.8571428571428571\\ 0.7857142857142857\\ 0.4\\ 0.8\\ 0.7\\ 0.5833333333333334\\ 0.9166666666666666\\ 0.6666666666666666\\  
	};
    \end{axis}
  \end{tikzpicture}
\caption[Immunity tasks completed]{Percentage of immunity tasks completed before deadline.}
\label{fig:mean-immunity-fixed-before-deadline}
\end{subfigure}

\vskip 1em

\begin{subfigure}[t]{0.90\columnwidth}
  \centering
  \begin{tikzpicture}
    \tikzstyle{every node}=[font=\small]
    \begin{axis}[
	y=0.7cm,
	ytick={1,2},
	yticklabel style={align=center},
	yticklabels={Individual,Group},
	width=7cm,
	xtick={0,20,40,60,80,100,120},
	xlabel={$\mu_{\mathrm{individual}} (23.12\%) <\mu_{\mathrm{group}} (46.84\%)$,\\$p=0.01$, $g$ = 0.75}, %0.0001
	xlabel style={align=center},
	xmin=-5, xmax=125,
	boxplot/average=auto,
	title style={align=center},
	title={},
	title style={yshift=-1ex,},
	]
	\addplot+[red,boxplot,mark options={fill=red}]
	table [row sep=\\,y index=0] {
	data\\
    % 0\\ 0\\ 0\\ 0\\ 0\\ 0\\ 0\\ 0\\ 117.6470588\\ 47.05882353\\ 40\\ 20\\ 0\\ 20\\ 12.5\\ 25\\ 12.5\\ 0\\ 100\\ 25\\ 50\\ 0\\ 16.66666667\\ 50\\ 50\\ 60\\ 20\\ 60\\ 0\\ 0\\ 72.72727273\\ 0\\ 0\\ 0\\ 90.90909091\\ 18.18181818\\ 0\\ 0\\ 44.44444444\\ 0\\ 0\\ 16.66666667\\ 0\\ 13.33333333\\ 85.71428571\\ 0\\ 0\\ 62.5\\ 25\\ 0\\
    0\\ 0\\ 0\\ 0\\ 0\\ 0\\ 0\\ 0\\ 117.6470588\\ 47.05882353\\ 40\\ 20\\ 0\\ 20\\ 12.5\\ 25\\ 12.5\\ 0\\ 100\\ 25\\ 50\\ 0\\ 16.66666667\\ 50\\ 50\\ 60\\ 20\\ 60\\ 0\\ 0\\ 72.72727273\\ 0\\ 0\\ 0\\ 90.90909091\\ 18.18181818\\ 0\\ 0\\ 44.44444444\\ 0\\ 0\\ 16.66666667\\ 0\\ 13.33333333\\ 85.71428571\\ 0\\ 0\\ 62.5\\ 25\\ 0\\     
	};
	\addplot+[blue,boxplot,mark options={fill=blue}]
	table [row sep=\\,y index=0] {
	data\\
	100\\ 100\\ 100\\ 100\\ 100\\ 44.44444444\\ 44.44444444\\ 44.44444444\\ 44.44444444\\ 44.44444444\\ 18.18181818\\ 18.18181818\\ 18.18181818\\ 18.18181818\\ 18.18181818\\ 44.44444444\\ 44.44444444\\ 44.44444444\\ 25\\ 25\\ 25\\ 61.53846154\\ 61.53846154\\ 61.53846154\\ 61.53846154\\ 66.66666667\\ 66.66666667\\ 66.66666667\\ 66.66666667\\ 60\\ 60\\ 60\\ 13.33333333\\ 13.33333333\\ 13.33333333\\ 100\\ 100\\ 100\\ 100\\ 71.42857143\\ 71.42857143\\ 71.42857143\\ 71.42857143\\ 0\\ 0\\ 0\\ 0\\ 0\\ 0\\ 0\\ 0\\ 60\\ 60\\ 60\\ 16.66666667\\ 16.66666667\\ 16.66666667\\
	};
    \end{axis}
  \end{tikzpicture}
\caption[Manager sanctions]{Number of manager sanctions per 100 attacks.}
\label{fig:mean-manager-sanction}
\end{subfigure}
% \sg{effect - 0.750276 , p - 0.000382454}

\vskip 1em

\begin{subfigure}[t]{0.90\columnwidth}
  \centering
  \begin{tikzpicture}
    \tikzstyle{every node}=[font=\small]
    \begin{axis}[
	y=0.7cm,
	ytick={1,2},
	yticklabel style={align=center},
	yticklabels={Individual,Group},
	width=7cm,
	xtick={-.6,-.3,0,.3,.6},
	xticklabels={$-60$, $-30$, $0$, $30$, $60$},
	xlabel={Percentage difference in immunity tasks completed \\ $\mu_{\mathrm{individual}} (13.24\%)>\mu_{\mathrm{group}} (0.68\%)$,\\$p=0.03$, $g$ = 0.70},
	xlabel style={align=center},
	xmin=-0.8, xmax=0.8,
	boxplot/average=auto,
	title style={align=center},
	title={},
	title style={yshift=-1ex,},
	]
	\addplot+[red,boxplot,mark options={fill=red}]
	table [row sep=\\,y index=0] {
	data\\
    0.0531135531136\\ 0.564705882353\\ 0.625882352941\\ 0.524705882353\\ 0.0258823529412\\ 0.225882352941\\ 0.195970695971\\ 0.148351648352\\ 0.166666666667\\ -0.266233766234\\ -0.121212121212\\ -0.266233766234\\ 0.0\\ -0.047619047619\\ 0.0952380952381\\ -0.0952380952381\\ 0.0808080808081\\ 0.313131313131\\ 0.17803030303\\ 0.0265151515152\\ 0.148148148148\\ 0.144249512671\\ 0.323586744639\\ 
	};
	\addplot+[blue,boxplot,mark options={fill=blue}]
	table [row sep=\\,y index=0] {
	data\\
	0.105555555556\\ 0.27962962963\\ -0.127777777778\\ 0.0\\ 0.0833333333333\\ 0.0\\ 0.0\\ 0.010582010582\\ 0.0952380952381\\ 0.042328042328\\ 0.0\\ -0.0416666666667\\ -0.047619047619\\ -0.0357142857143\\ -0.200483091787\\ -0.200483091787\\ -0.0214285714286\\ -0.0190476190476\\ 0.109848484848\\ 0.094696969697\\ 0.0151515151515\\ 
	};
    \end{axis}
  \end{tikzpicture}
\caption[Immunity tasks first and second game]{Difference in immunity tasks completed before deadline between first game (InS$_1$ or GrS$_1$) and second game (InS$_2$ or GrS$_2$). }

\label{fig:difference-immunities}
\end{subfigure}
% \sg{effect - .710025, p - 0.020052203}

\vskip 1em

\begin{subfigure}[t]{0.90\columnwidth}
\centering
  \begin{tikzpicture}
     \tikzstyle{every node}=[font=\small]
     \begin{axis}[
 	y=0.7cm,
 	ytick={1,2,3,4},
 	yticklabel style={align=center},
 	yticklabels={},
 	width=7cm,
 	xlabel={Immunity tasks completed before deadline \\$\mu$\fsub{I\fsub{RA}} (50.36\%) $<$ $\mu$\fsub{I\fsub{RS}} (66.51\%); $p < 0.01$.},
 	xlabel style={align=center},
 	boxplot/average=auto,
 	title style={align=center},
 	title={},
 	title style={yshift=-1ex,},
 	ytick={1,2,3},
 	yticklabels={Risk averse, Risk seeking},
 	]
%  	\addplot+[magenta,boxplot,mark options={fill=magenta!50!black}]
%  	table[y=number_of_manager_sanctions, col sep=comma]
%  	{data/I-all-risk-averse.csv};
%  	\addplot+[olive,boxplot,mark options={fill=olive!50!black}]
%  	table[y=number_of_manager_sanctions, col sep=comma]
%  	{data/I-all-risk-n+t.csv};
 	\addplot+[magenta,boxplot,mark options={fill=magenta!50!black}]
 	table[y=immunity_ratio_completed_available, col sep=comma]
 	{data/all-risk-averse.csv};
 	\addplot+[olive,boxplot,mark options={fill=olive!50!black}]
 	table[y=immunity_ratio_completed_available, col sep=comma]
 	{data/all-risk-n+t.csv}; 	
     \end{axis}
\end{tikzpicture}
\caption{Risk-seeking (RS) players completed significantly more immunity tasks before deadline than risk-averse (RA) players. }
\label{fig:risk-immunity-deadline}
\end{subfigure}

\caption{Influence of sanctions on compliance.}
\label{fig:compliance}
\end{figure}

 Figure~\ref{fig:mean-immunity-fixed-before-deadline} shows the percentage of immunity tasks completed before their deadline. Individual sanction offers a small improvement in compliance over group sanction.

We calculate the number of manager sanctions per number of attacks in a game. Figure~\ref{fig:mean-manager-sanction} shows that on average, for every 100 attacks, there were 23 instances of players being sanctioned under individual sanction and 47 instances of players being sanctioned under group sanction. This medium effect is statistically significant to 1\% but compatible with the result in Figure~\ref{fig:mean-immunity-fixed-before-deadline}. 

For each sanction type, we calculated the difference between the number of immunity tasks per attack completed by a player to measure the influence of sanction on the player. Figure~\ref{fig:difference-immunities} shows that on average a player completed 13\% more immunity tasks per attack in the second game (InS$_2$)
than in the first game (InS$_1$) with individual sanctions. For group sanctions, the number of immunity tasks completed in games GrS$_1$ and GrS$_2$ was unaffected. This medium effect is statistically significant to 2\%, signifying that individual sanction was more effective than group sanction in motivating people to be security compliant. 
We conjecture that whereas (1) experience improves the advantage of individual sanctions in promoting compliance, (2) experience has no effect for group sanctions.

We found 14 instances of peer sanction in games with group sanction ($\mu=1.24$) and seven instances of peer sanction in games with individual sanction ($\mu=0.75$).  This indicates that group sanctions promote peer sanctions more than individual sanctions.

To investigate if risk attitudes influence compliance, we categorize each player as a \emph{risk-seeking} player or a \emph{risk-averse} player based on the players' social and financial risk scores obtained from the DOSPERT survey. 

Figure~\ref{fig:risk-immunity-deadline} compares the percentage of immunity tasks completed before their deadline by risk-seeking and risk-averse players. We find that risk-seeking players are more compliant---complete more immunity tasks by the deadline---than risk-averse players, and as a result, are sanctioned less often. We observe this trend---where risk-seeking players complete the immunity tasks though near the deadline---with both individual and group sanction games. 

After each game, players rated on a Likert scale of 1 (\fsl{not at all influential}) to 5 (\fsl{very influential}), how effective the sanctioning mechanism was. We found that 77 percent of participants identified sanctions as a strong factor (4--5) in influencing their decision making.

\subsection{Efficacy}

Figure~\ref{fig:mean-score} shows the scores earned by the players. Participants were more productive in individual sanction games. The average score in individual sanction games (373) was greater than in group sanction games (334) with a small effect size but was statistically significant to 1\%. The result is not surprising because with group sanctions, compliant players were sanctioned because of noncompliance by others in their group. 

\begin{figure}[!htb]
    \centering
\begin{subfigure}[t]{0.90\columnwidth}
  \centering
  \begin{tikzpicture}
    \tikzstyle{every node}=[font=\small]
    \begin{axis}[
	y=0.7cm,
	ytick={1,2},
	yticklabel style={align=center},
	yticklabels={Individual,Group},
	width=7cm,
    xtick={100, 200, 300, 400, 500},
	xlabel={$\mu_{\mathrm{individual}}>\mu_{\mathrm{group}}$, $p=0.01$, $g$ = 0.45}, %0.0152
	xlabel style={align=center},
	xmin=50, xmax=550,
	boxplot/average=auto,
	title style={align=center},
	title={},
	title style={yshift=-1ex,},
	]
	\addplot+[red,boxplot,mark options={fill=red}]
	table [row sep=\\,y index=0] {
	data\\
   540\\ 470\\ 530\\ 495\\ 435\\ 270\\ 225\\ 270\\ 190\\ 160\\ 385\\ 270\\ 405\\ 415\\ 315\\ 190\\ 315\\ 360\\ 375\\ 370\\ 335\\ 380\\ 315\\ 305\\ 340\\ 460\\ 450\\ 450\\ 430\\ 385\\ 280\\ 450\\ 405\\ 450\\ 345\\ 450\\ 440\\ 410\\ 405\\ 405\\ 350\\ 400\\ 315\\ 370\\ 340\\ 495\\ 450\\ 315\\ 360\\ 355\\
	};
	\addplot+[blue,boxplot,mark options={fill=blue}]
	table [row sep=\\,y index=0] {
	data\\
	210\\ 225\\ 185\\ 180\\ 120\\ 395\\ 360\\ 360\\ 395\\ 235\\ 385\\ 340\\ 385\\ 380\\ 295\\ 400\\ 370\\ 405\\ 235\\ 405\\ 270\\ 245\\ 245\\ 330\\ 210\\ 345\\ 290\\ 330\\ 225\\ 415\\ 405\\ 430\\ 470\\ 405\\ 430\\ 315\\ 315\\ 335\\ 315\\ 230\\ 225\\ 200\\ 235\\ 405\\ 415\\ 495\\ 480\\ 405\\ 375\\ 405\\ 405\\ 340\\ 335\\ 280\\ 440\\ 405\\ 365\\
	};
    \end{axis}
  \end{tikzpicture}
\mycaption{Score of players in a game}
\label{fig:mean-score}
\end{subfigure}
% \sg{effect - 0.453226, p - 0.013698625}
% 
% \pgfplotstableread[row sep=\\,col sep=&]{
%      	quality & Breach Type\\
% 			7.64 & Individual\\
% 			5.421052632 & Group\\
%  }\averagetimetofiximmunity
% 

\vskip 1em

\begin{subfigure}[t]{0.90\columnwidth}
  \centering
  \begin{tikzpicture}
    \tikzstyle{every node}=[font=\small]
    \begin{axis}[
	y=0.7cm,
	ytick={1,2},
	yticklabel style={align=center},
	yticklabels={Individual,Group},
	width=7cm,
	xtick={0,4,8,12, 16},
	xlabel={$\mu_{\mathrm{individual}}<\mu_{\mathrm{group}}$, $p<0.01$, $g$ = 0.82}, %0.0002
	xlabel style={align=center},
	xmin=-1, xmax=18,
	boxplot/average=auto,
	title style={align=center},
	title={},
	title style={yshift=-1ex,},
	]
	\addplot+[red,boxplot,mark options={fill=red}]
	table [row sep=\\,y index=0] {
	data\\
    0\\ 0\\ 0\\ 0\\ 0\\ 0\\ 0\\ 0\\ 17\\ 8\\ 4\\ 2\\ 0\\ 2\\ 2\\ 4\\ 2\\ 0\\ 8\\ 2\\ 4\\ 3\\ 4\\ 2\\ 4\\ 5\\ 2\\ 6\\ 0\\ 0\\ 8\\ 0\\ 0\\ 0\\ 10\\ 2\\ 0\\ 0\\ 4\\ 0\\ 0\\ 2\\ 0\\ 2\\ 6\\ 2\\ 1\\ 7\\ 4\\ 1\\
	};
	\addplot+[blue,boxplot,mark options={fill=blue}]
	table [row sep=\\,y index=0] {
	data\\
	14\\ 14\\ 14\\ 14\\ 14\\ 4\\ 4\\ 4\\ 4\\ 4\\ 2\\ 2\\ 3\\ 3\\ 2\\ 4\\ 4\\ 4\\ 4\\ 4\\ 4\\ 13\\ 10\\ 8\\ 9\\ 9\\ 8\\ 9\\ 9\\ 5\\ 5\\ 5\\ 2\\ 2\\ 2\\ 10\\ 10\\ 10\\ 10\\ 11\\ 11\\ 12\\ 12\\ 0\\ 0\\ 0\\ 0\\ 0\\ 0\\ 0\\ 0\\ 6\\ 6\\ 6\\ 2\\ 2\\ 2\\
	};
    \end{axis}
  \end{tikzpicture}
\caption[Rounds passed]{Number of rounds skipped.}
\label{fig:mean-rounds-passed}
\end{subfigure}

\begin{subfigure}[t]{0.90\columnwidth}
  \centering
\begin{tikzpicture}
     \tikzstyle{every node}=[font=\small]
     \begin{axis}[
 	y=0.7cm,
 	ytick={1,2,3,4},
 	yticklabel style={align=center},
 	yticklabels={},
 	width=7cm,
 	xlabel={Score per task\\$\mu$\fsub{RS} (13.54) $>$ $\mu$\fsub{RS} (12.12); $p < 0.01$.},
 	xlabel style={align=center},
 	boxplot/average=auto,
 	title style={align=center},
 	title={},
 	title style={yshift=-1ex,},
 	ytick={1,2,3},
 	yticklabels={Risk averse, Risk seeking},
 	]
 	\addplot+[magenta,boxplot,mark options={fill=magenta!50!black}]
 	table[y=per_task_point, col sep=comma]
 	{data/all-risk-averse.csv};
 	\addplot+[olive,boxplot,mark options={fill=olive!50!black}]
 	table[y=per_task_point, col sep=comma]
 	{data/all-risk-n+t.csv};
     \end{axis}
\end{tikzpicture}
\caption{Risk seekers (RS) prefer completing higher valued tasks than risk-averse participants (RA).}
\label{fig:risk-task-scores}
\end{subfigure}

\caption{Influence of sanctions and risk attitudes on productivity.}
\label{fig:productivity}
\end{figure}
% \sg{effect - 0.824402 , p - 0.00012915}

Figure~\ref{fig:mean-rounds-passed} shows the number of rounds a player skipped in a game. 
On average, the number of rounds passed in group sanction games is double of those passed in the individual sanction games, indicating a large effect at the 1\% significance. This observation on rounds passed explains how group sanctions produce lower scores than individual sanctions, as Figure~\ref{fig:mean-score} shows. 

Figure~\ref{fig:risk-task-scores} shows the average score gained per task. 
We observe that risk-seeking participants complete more high valued tasks than risk-averse players. This observation is equally more prominent under individual sanction games than group sanction games.

After each game, we asked participants to rate how detrimental the sanctioning mechanism was to completion of tasks---on a Likert scale of 1 to 5 with 1 being \fsl{very detrimental} and 5 being \fsl{not at all detrimental}. 
We find that participants perceive both group sanctions and individual sanctions as equally detrimental to their productivity. 

\subsection{Resilience}

\begin{figure}[!htb] 
  \centering
\begin{subfigure}[t]{0.90\columnwidth}
  \centering
  
  \begin{tikzpicture}
    \tikzstyle{every node}=[font=\small]
    \begin{axis}[
	y=0.7cm,
	ytick={1,2},
	yticklabel style={align=center},
	yticklabels={Individual,Group},
	width=7cm,
	xlabel={$\mu_{\mathrm{individual}}<\mu_{\mathrm{group}}$, $p=0.11$},%, $g$ = 0.29}, %0.0004
	xlabel style={align=center},
	boxplot/average=auto,
	title style={align=center},
	title={},
	title style={yshift=-1ex,},
	]
 	\addplot+[red,boxplot,mark options={fill=red}]
 	table[y=time_per_immunity, col sep=comma]
 	{data/G-all.csv};
 	\addplot+[blue,boxplot,mark options={fill=blue}]
 	table[y=time_per_immunity, col sep=comma]
 	{data/I-all.csv};

    \end{axis}
  \end{tikzpicture}
\caption[Resilience]{Comparing sanction types on rounds taken to regain immunity, indicating resilience.}
\label{fig:time-taken-to-regain-immunity}
\end{subfigure}
\begin{subfigure}[t]{0.90\columnwidth}
  \centering
  \begin{tikzpicture}
     \tikzstyle{every node}=[font=\small]
     \begin{axis}[
 	y=0.7cm,
 	ytick={1,2,3,4},
 	yticklabel style={align=center},
 	yticklabels={},
 	width=7cm,
 	xlabel={Time (in rounds) to regain immunity \\ $\mu$\fsub{RS} (0.56) $>$ $\mu$\fsub{RA} (0.36); $p < 0.01$.},
 	xlabel style={align=center},
 	boxplot/average=auto,
 	title style={align=center},
 	title={},
 	title style={yshift=-1ex,},
 	ytick={1,2,3},
 	yticklabels={Risk averse, Risk seeking},
 	]
 	\addplot+[magenta,boxplot,mark options={fill=magenta!50!black}]
%  	table[y=time_to_fix_immunity, col sep=comma]
    table[y=time_per_immunity, col sep=comma]
 	{data/all-risk-averse.csv};
 	\addplot+[olive,boxplot,mark options={fill=olive!50!black}]
%  	table[y=time_to_fix_immunity, col sep=comma]
    table[y=time_per_immunity, col sep=comma]
 	{data/all-risk-n+t.csv};
     \end{axis}
\end{tikzpicture}
\caption{Risk seekers (RS) wait significantly longer to complete immunity tasks than risk-averse (RA) participants.}
\label{fig:risk-immunity-regain}  
\end{subfigure}
\caption{Influence of sanctions and risk attitudes on resilience.}
\end{figure}
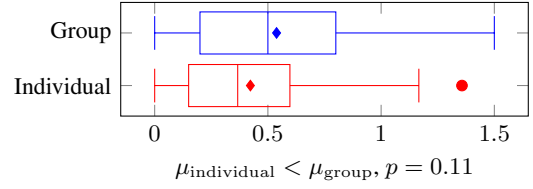
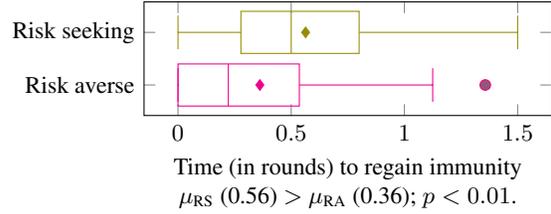

We calculated the number of rounds a player took to regain immunity lost to an attack. 
Figure~\ref{fig:time-taken-to-regain-immunity} shows that whereas both group sanction and individual sanction motivated participants to complete the immunity task immediately (both have a mean close to zero), individual sanction sees a small improvement in resilience over group sanction. 

Although we observe no significant difference in resilience offered by the two sanction types, we notice (Figure~\ref{fig:risk-immunity-regain}) that risk seekers wait significantly longer to regain immunity after losing it to an attack. On further analysis, we find that this difference is prominent only with individual sanctions. Under group sanctioning, there is no significant difference in rounds players took to regain an immunity. 
We conjecture that group sanctioning could be more effective and could offer better resilience in an organization where a majority of workers are risk seekers. 

\section{Conclusions}
\label{sec:concl}
We investigate the effectiveness of group, individual, and peer sanctions in promoting cybersecurity while recognizing the overhead potentially caused by cybersecurity activities on productivity. Our empirical study suggests that individual sanctions are more effective in enforcing compliance with cybersecurity regulations than group sanctions. 
 
Workers are less productive in completing project tasks under group sanction than under individual sanction. Interestingly, the presence of group sanction leads to increased peer sanctions, indicating the potential for obtaining a self-sustaining system. 

Sanctions, when leading to loss of work or acting as blockers in completing daily task, were more effective for risk-averse workers than risk seekers. Overall, individual sanctioning is more effective than group sanctioning in realizing cybersecurity practices.

\subsection{Threats to Validity and Mitigation}
First, players may not be motivated to give their best. Hence their game moves may not be carefully thought out.
We offered a bonus based on game performance. Second, to ensure the game is understood well, the players watched a video explaining the game in detail and played two demo games. Third, as with any study, there is a risk that the participants may not be representative of the larger population.

\subsection{Future Directions}
First, experiments combining group and peer sanction with the observability of noncompliance being variable would be interesting. Monitoring can be decreased gradually, relying increasingly on peer sanction to keep workers security compliant. This type of system would scale better, but the effect of this on the productivity of participants is yet to be seen. 

Second, it would be interesting to explore positive sanctions in future studies where complying with security regulations is rewarded. The final score could be based on a weighted average of the numbers of security and project tasks completed, not just the project tasks, as currently.

%%%%%%%%%%%%%%%%%%%%%%%%%%%%%%%%%%%%%%%%%%%%%%%%%%%%%%
\section*{Acknowledgments}
%%%%%%%%%%%%%%%%%%%%%%%%%%%%%%%%%%%%%%%%%%%%%%%%%%%%%%
This research was supported by the Science of Security Lablet at NC State University. 

% %%%%%%%%%%%%%%%%%%%%%%%%%%%%%%%%%%%%%%%%%%%%%%%%%%%%%%
% \section*{Author Bios}
% %%%%%%%%%%%%%%%%%%%%%%%%%%%%%%%%%%%%%%%%%%%%%%%%%%%%%%
% \begin{description}
% \item [Nirav Ajmeri] is a Postdoctoral Research Scholar in Computer Science at NC State University. 
% His research interests include artificial intelligence and multiagent systems with a focus on cybersecurity and privacy. 
% % His research seeks to facilitate engineering a society of socially intelligent agents that act ethically and yield a satisfactory experience to their users. 
% Nirav received a PhD in Computer Science from NC State University. Contact him at najmeri@ncsu.edu.

% \item [Shubham Goyal] is a software developer at Amazon. He received an MS in Computer Science from NC State University. 
% Contact him at gsshubha@ncsu.edu.

% \item [Munindar P.~Singh] is a Professor in Computer Science and a co-director of the Science of Security Lablet at NC State University.  His research interests include the engineering and governance of sociotechnical systems.  Singh is an IEEE Fellow, a AAAI fellow, and a former Editor-in-Chief of \emph{IEEE Internet Computing} and the \emph{ACM Transactions on Internet Technology}. Contact him at singh@ncsu.edu.

% \end{description}

%%%%%%%%%%%%%%
% \clearpage
\bibliographystyle{unsrtnat}
\bibliography{Munindar,Nirav,Shubham}
\balance

\end{document}